\documentclass[]{JHEP3}
\usepackage{graphicx}
\JHEPspecialurl{http://jhep.sissa.it/JOURNAL/JHEP3.tar.gz}

\usepackage{amsmath}
\usepackage{epsfig,multicol,bbm}
\usepackage{url}

\newcommand{\newc}{\newcommand}
\newc{\gsim}{\lower.7ex\hbox{$\;\stackrel{\textstyle>}{\sim}\;$}}
\newc{\lsim}{\lower.7ex\hbox{$\;\stackrel{\textstyle<}{\sim}\;$}}
\newc{\gev}{\,{\rm GeV}}
\newc{\mev}{\,{\rm MeV}}
\newc{\ev}{\,{\rm eV}}
\newc{\kev}{\,{\rm keV}}
\newc{\tev}{\,{\rm TeV}}
\def\ln{\mathop{\rm ln}}

\def\Tr{\mathop{\rm Tr}}

\newc{\mz}{M_Z}
\newc{\mpl}{M_*}
\newc{\mw}{m_{\rm weak}}
\newc{\nr}[1]{N^c_R{}_{#1}}
\usepackage{amsmath}
\def\beq{\begin{equation}}
\def\eeq{\end{equation}}
\def\bea{\begin{eqnarray}}
\def\eea{\end{eqnarray}}
\def\bitem{\begin{itemize}}
\def\eitem{\end{itemize}}
\newc{\ie}{{\it i.e.}}          \newc{\etal}{{\it et al.}}
\newc{\eg}{{\it e.g.}}          \newc{\etc}{{\it etc.}}
\newc{\cf}{{\it c.f.}}
\newcommand{\kahler}{K\"{a}hler }

\newcommand{\lang}{\mathcal{L}}

\newcommand{\half}{\frac{1}{2}}

\def\vev#1{\left\langle #1 \right\rangle}

\def\inv{^{\raise.15ex\hbox{${\scriptscriptstyle -}$}\kern-.05em 1}}
\def\lbar{{\lower.35ex\hbox{$\mathchar'26$}\mkern-10mu\lambda}} 

\def\to{\rightarrow}

\let\<=\langle
\let\>=\rangle

\let\+=\uparrow

\newcommand\fverb{\setbox\fverbbox=\hbox\bgroup\verb}
\newcommand\fverbdo{\egroup\medskip\noindent%
			\fbox{\unhbox\fverbbox}\ }
\newcommand\fverbit{\egroup\item[\fbox{\unhbox\fverbbox}]}
\newbox\fverbbox

\title{The Goldstini Variations}

\date{\today}

\author{Nathaniel Craig\\
	Institute for Theoretical Physics, Stanford University, Stanford, CA 94306, USA\\
	E-mail: \email{ncraig@stanford.edu}}

\author{John March-Russell\\
	Rudolf Peierls Centre for Theoretical Physics, University of Oxford, 1 Keble Road, Oxford, OX1 3NP, UK\\
	E-mail: \email{jmr@thphys.ox.ac.uk}}

\author{Matthew McCullough\\
	Rudolf Peierls Centre for Theoretical Physics, University of Oxford, 1 Keble Road, Oxford, OX1 3NP, UK\\
	E-mail: \email{mccull@thphys.ox.ac.uk}}

\preprint{OUTP-10-11P \\ SU-ITP-10-23}	

\abstract{We study the `goldstini' scenario of Cheung, Nomura, and Thaler, in which multiple independent
supersymmetry (SUSY) breaking sectors lead to multiple would-be goldstinos, changing collider and cosmological phenomenology.  In supergravity, potentially large corrections to the previous $2 m_{3/2}$ prediction for goldstini masses can arise when their scalar partners are stabilized far from the origin.  Considerations arising from the complexity of realistic string compactifications indicate that many of the independent SUSY-breaking sectors should be conformally sequestered or situated in warped Randall-Sundrum-like throats, further changing the predicted goldstini masses.  If the sequestered hidden sector is a metastable SUSY-breaking sector of the Intriligator-Seiberg-Shih (ISS) type then multiple goldstini can originate from within a single sector, along with many supplementary `modulini', all with masses of order $2 m_{3/2}$.  These fields can couple to the Supersymmetric Standard Model (SSM) via the `Goldstino Portal'.   Collider signatures involving SSM sparticle decays can provide strong evidence for warped-or-conformally-sequestered sectors, and of the ISS mechanism of SUSY breaking.   Along with axions and photini, the Goldstino Portal gives another potential window to the hidden sectors of string theory.}


\begin{document} 

\maketitle

\section{Introduction}

If the Standard Model is UV completed by string theory -- consistent with the hypothesis of supersymmetry (SUSY) -- the topological complexity of realistic compactification manifolds suggests the existence of many additional sectors sequestered from the fields of the Standard Model.  The dimensional reduction of form fields may result in a proliferation of light axion-like scalar fields \cite{Arvanitaki:2009fg,Arvanitaki:2010sy}, or weak-scale abelian vector fields and their superpartners \cite{Arvanitaki:2009hb}, which can dramatically alter standard cosmological, astrophysical, and collider phenomenology.  Moreover, the presence of stacks of spacetime-filling branes may lead to nonabelian gauge sectors with fundamental matter in the four-dimensional theory.  The mere observation that such supersymmetric nonabelian gauge theories possess metastable SUSY-breaking vacua \cite{ISS} suggests that supersymmetry may be broken in these different (purely field-theoretic) sectors. Furthermore, there are numerous additional ways in which supersymmetry may be broken by intrinsically stringy objects -- e.g., nonsupersymmetric flux backgrounds or the presence of both D- and anti D-branes in the compactification manifold.  On a topologically complex compactification manifold with various nonabelian gauge sectors, fluxes, branes, and antibranes, it is not unreasonable to expect a rich variety of SUSY-breaking dynamics to coexist.  Thus, the existence of multiple (likely metastable) SUSY-breaking sectors is not merely a theoretical novelty, but rather a well-motivated consequence of physics in the ultraviolet.\footnote{Indeed, the fact that cosmological evolution preferentially populates the metastable vacua of SQCD rather than the supersymmetric vacua \cite{Abel:2006cr, Craig:2006kx, Fischler:2006xh} provides a strong argument that the mere existence of multiple (reheated) nonabelian gauge sectors with light fundamental matter implies the existence of simultaneous SUSY breaking in multiple sectors.}   

Historically, however, the study of SUSY breaking and its phenomenology has focused on a single sector additional to the Supersymmetric Standard Model (SSM), whose dynamics give rise to a nonsupersymmetric ground state.  Recently it has been shown \cite{Benakli:2007zza,Goldstini1,Goldstini2} that relaxing this assumption to include multiple sources of SUSY breaking can lead to interesting and appealing scenarios in which the conventional phenomenology of single-sector SUSY breaking is significantly modified, similar to the way in which multiple photini can also alter conventional SUSY phenomenology \cite{Arvanitaki:2009hb}. In this paper we wish to extend the results of \cite{Goldstini1,Goldstini2} with an eye towards the underlying physical context in which multiple SUSY breaking is likely to arise.

The mediation of this multiple-sector supersymmetry breaking to the Standard Model may occur in any of the customary ways, leading to weak-scale soft masses and the usual successes of the SSM. However, even if the fundamental interactions between these sectors and the SSM are Planck-suppressed, the multiple breaking of supersymmetry gives rise to less-suppressed couplings between additional `goldstini' and SSM fields. In this fashion, the existence of new sectors with otherwise-unobservable couplings to the SSM may be revealed via what
we may call the `Goldstino Portal'.  In this sense the goldstini and their companions are further distinguished from moduli, whose masses are likewise around $m_{3/2}$ but whose couplings are Planck-suppressed.

Specifically, in \cite{Benakli:2007zza,Goldstini1} it was argued that the presence of multiple sequestered sectors that break SUSY spontaneously gives rise to multiple `goldstini' in addition to the true global goldstino which provides the extra degrees of freedom of the gravitino of mass $m_{3/2}$, and in \cite{Goldstini1} it was shown that that these additional goldstini would have mass $2 m_{3/2}$.   It was further shown \cite{Goldstini1,Goldstini2} that such a set-up can lead to exciting new signatures at the LHC which could confirm not only the validity of the supergravity framework but also the presence of multiple sequestered SUSY breaking sectors, providing indirect, but striking, evidence for complexity of the string compactification.  Subsequently it was shown that this scenario can also lead to solutions of the cosmological problems with a heavy gravitino LSP \cite{Goldstini2}. 

These considerations come with two caveats. The first is purely experimental; the smallness of flavour-changing neutral currents (FCNCs) and other signs of Standard Model flavour violation imply that the flavour-violating contributions to SSM soft masses from all SUSY-breaking sectors must necessarily be small.  In particular, this requires that all SUSY breaking communicated via intrinsically flavour-violating mediation mechanisms such as (the non-anomaly-mediated part of) gravity mediation must be at most one thousand times smaller than that communicated via flavour-preserving mechanisms.  Although this is possible if all such contributions to SUSY breaking are conveniently small to begin with, it seems much more plausible that the smallness of flavor violation arises from locality and warping \cite{Randall:1998uk} or conformal sequestering \cite{Luty:2001jh, Luty:2001zv}.
Once again, this is a well-motivated consequence of physics in the ultraviolet.  Sequestering is known to arise readily in the presence of strongly warped backgrounds such as warped throats, e.g., type IIB string theory \cite{Kachru:2007xp}, and the ubiquity of warped throats on realistic compactification manifolds is well-known \cite{Giddings:2001yu,Dimopoulos:2001qd,Hebecker:2006bn}. The pairing of multiple SUSY breaking and sequestering via warped throats is suggested by more than just FCNC considerations alone; the very existence of multiple goldstini requires it as multiple {\it unsequestered} SUSY breaking sectors simply lead to one ur-breaking of supersymmetry.  But if sequestering and multiple SUSY breaking are so closely intertwined, it is then natural to consider what implications sequestering may have on the spectrum and phenomenology of the resulting goldstini.  In particular, we will argue below that warping and sequestering lead to substantial deviations from the goldstino mass prediction of $2 m_{3/2}$, and that the spectrum of goldstini -- and resulting collider phenomenology -- are  richer than previously thought.

The second consideration is largely theoretical.  Weak-scale supersymmetry in the SSM favours {\it dynamical} means of SUSY breaking in order to explain the hierarchy between the Planck and SUSY-breaking scales \cite{Witten:1981nf}.  In turn, dynamical SUSY breaking in general requires a SUSY-breaking sector to possess a rich set of gauge dynamics and fields. It is therefore instrumental to consider whether common classes of dynamical SUSY-breaking theories might modify or alter the goldstino spectrum, perhaps by the presence of additional light states.  At the very least, the corresponding goldstino mass depends on how the SUSY breaking vacuum is stabilized.  Moreover, we will argue that it is quite common that a {\it single} dynamical SUSY-breaking sector gives rise to {\it multiple faux goldstini}.  Such additional states may then couple to MSSM fields through the Goldstino Portal, and their observation would shed further light on the nature of the supersymmetry breaking sector(s). 

In short: the potential observability of multiple SUSY breaking has been well established. However, it is instrumental to ask whether the additional physics that naturally accompanies multiple SUSY breaking may enrich and expand the goldstino spectrum and phenomenology.

\begin{figure}[]
\centering
\includegraphics[height=2.0in]{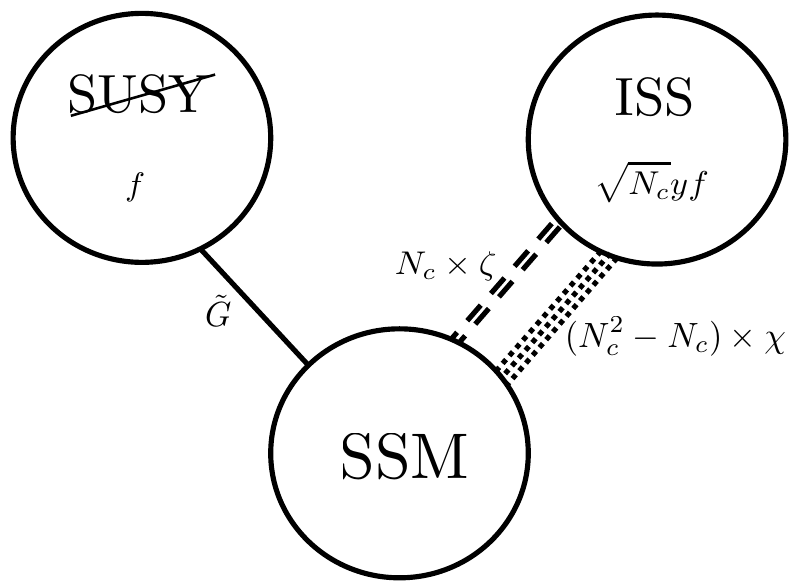}\caption{A schematic example of multiple sequestered SUSY-breaking.  In this setup there are two SUSY breaking sectors: The first sector is a SUSY breaking sector with one F-term of magnitude $f$ which couples to the SSM with an effective mediation scale $\Lambda$.  This sector needn't preserve an R-symmetry and could thus generate gaugino masses.  The second sector is an R-symmetry preserving $SU(N_c)$ ISS sector where all non-zero F-terms are of magnitude $y f$, implying this sector has an overall effective-SUSY-breaking-scale of $\sqrt{N_c} y f$.  The ISS sector couples to the SSM with a mediation scale of $\sqrt{x} \Lambda$.  The overall effective SUSY breaking scale that determines the gravitino mass is $f_{eff} = f \sqrt{1 + N_c y^2}$.  If $y\ll 1$, $N_c$ goldstini, $\zeta$, and $N_c(N_c-1)$ modulini, $\chi$, all arise from the ISS sector, as shown in Section~4, while the longitudinal mode of the gravitino dominantly arises from the first sector.}
\label{fig:example}
\end{figure}

In particular, in Section \ref{mass} we compute the goldstino mass for a general class of effective supersymmetry breaking Lagrangians using the conformal compensator formalism.  We will find that important corrections to the goldstino mass arise from the effects related to the stabilization of the SUSY-breaking vacuum. In Section \ref{warpandseq} we show how the presence of warping, or conformal sequestering, significantly modifies the prediction of $2 m_{3/2}$ for the goldstini masses.  The discovery of such modifications would be `smoking gun' for  the presence of such dynamics in one or more sequestered hidden sectors.  In Section \ref{ISS} we study the particle content of a hidden Intriligator-Seiberg-Shih-type (ISS) \cite{ISS} sector preserving a (discrete) R-symmetry.   We show that such a sector would give rise to $N_c$ goldstini and $N_c (N_c-1)$ `modulini' of mass $\geq 2 m_{3/2}$ (in the absence of warping or conformal sequestering), where $N_c$ is the number of colours in the asymptotically free UV gauge group.  A simple example of this setup is schematically illustrated in Figure~\ref{fig:example}.  
Although such a discrete-R-symmetry-preserving sector is incapable of generating gaugino masses, in the context of multiple SUSY breaking sectors this poses no problem.  In particular, since there is no phenomenological reason to require more than one of the independent SUSY-breaking sectors to break R-symmetry, and requiring all independent sectors to break R-symmetry severely constrains the number of possible string landscape vacua, we expect that our results on ISS-type sectors should apply to realistic theories of SUSY-breaking in the string landscape.\footnote{We note in passing that such R-symmetry-preserving sequestered SUSY breaking sectors can also lead to attractive phenomenological features, such as, \eg, cosmologically acceptable thermal leptogenesis \cite{Goldstini2}.}

\begin{figure}[]
\centering
\includegraphics[height=2.34in]{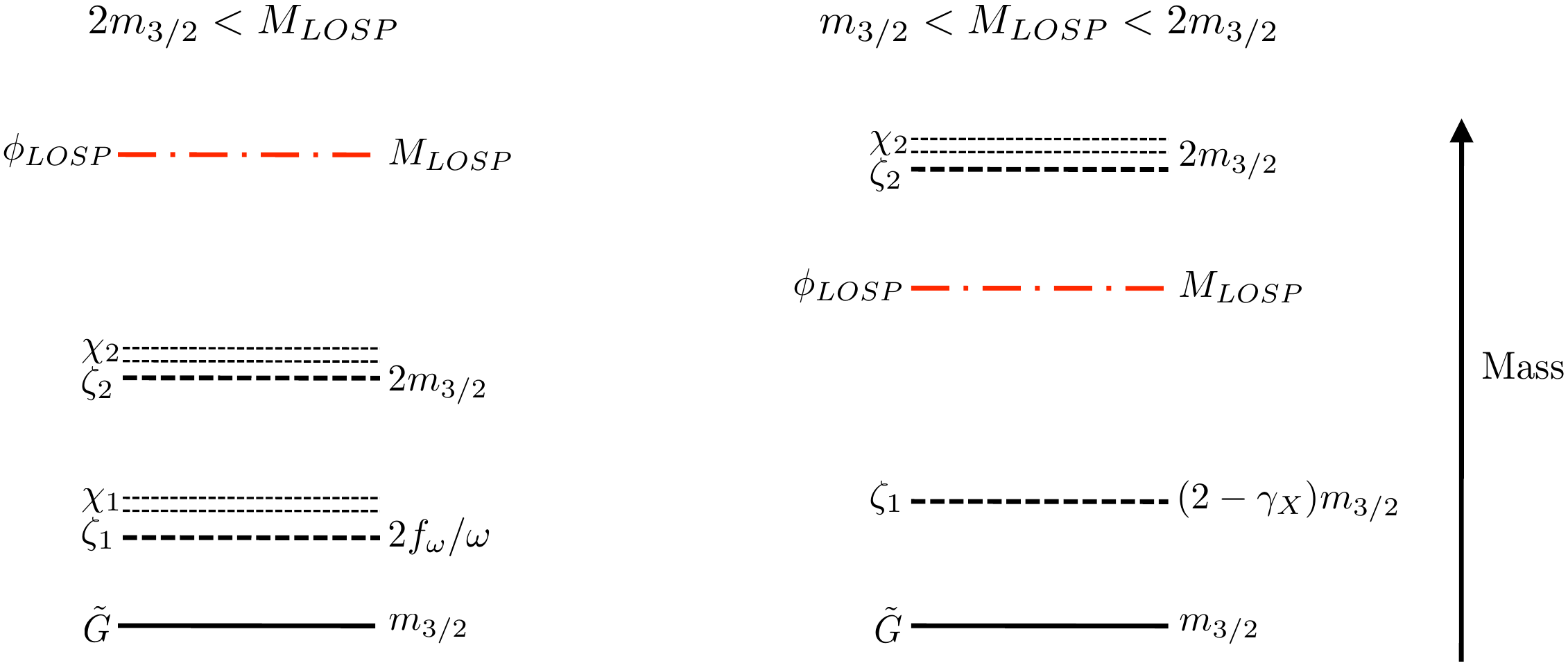}\caption{A diagram depicting a subset of the possible spectra.  The left panel shows the SSM LOSP, the gravitino and goldstini/modulini from two ISS sectors, one at the end of a warped throat (so with mass spectrum at
$2 f_\omega/\omega$ as shown in Section~3.1), and one just gravitationally sequestered from the SSM.  The right panel shows a possible spectrum where the SSM LOSP is lighter than $2 m_{3/2}$, but, however, could still decay to a goldstino originating in a conformally sequestered (or warped) sector, here chosen not to be of ISS type, so there is only a single goldstini state, and no modulini.  An interesting variant of this scenario occurs if the anomalous dimension of the SUSY-breaking field satisfies $\gamma_X > 1$, in which case LOSP decays could occur to a goldstino which is lighter than the gravitino.  The resulting collider and cosmological phenomenology can depend strongly on which of these patterns is realised.  Unlike for the various goldstini, decays to the modulini of a hidden ISS sector depend on the couplings in that sector, and are thus not guaranteed.  Three different sectors are shown in order to elucidate a range of possibilities, though any number $\geq 2$ of independent SUSY-breaking sectors implements the goldstini scenario.}
\label{fig:possibilities}
\end{figure}

Figure \ref{fig:possibilities} provides a schematic illustration of the new possibilities that arise for the mass spectrum of goldstini/modulini. 
Such states are typically grouped into sets with a goldstino (or goldstini) at a lower limit point at $\beta m_{3/2}$ with modulini sitting
relatively tightly spaced above this limit.  For unsequestered or unwarped sectors, $\beta=2$ (modulo potentially large corrections related to the stabilization of the SUSY-breaking vacuum as explained in Section \ref{mass}).  Otherwise, any value $2\geq \beta\geq 0$
is possible, so some subset of the goldstini/modulini may be lighter then the gravitino, while the lightest observable-sector supersymmetric partner (LOSP) may either sit above all the goldstini and modulini, or may be in the middle of the spectrum of states.  We emphasise
that in theories with multiple sectors it is unreasonable to expect, indeed unlikely, for the LSP to reside in our sector.\footnote{We particularly
thank Lawrence Hall for stressing the importance of this point to us.}  The true LSP may
be the gravitino, one of the limit point goldstini, or yet another state, such as a hidden photino.

In Section \ref{pheno} we briefly discuss the couplings of the goldstini and modulini of our secnario, and among other topics, present a potential `smoking gun' collider signature that can give evidence for the physical realisation of the ISS mechanism of SUSY-breaking.\footnote{We will return to the detailed phenomenology of the goldstini/modulini and their effects on collider experiments and astrophysical and cosmological observations in a later work.}  
In general the goldstini of multiple SUSY breaking sectors, including those within a hidden ISS sector, couple to SSM chiral multiplets through the Goldstino Portal as \cite{Goldstini1}
\beq
\mathcal{L}_{int} \supset \sum_{i=1}^{N} \sum_{a=1}^{N-1} \frac{\tilde{m}^2_i V_{ia}}{f_i} \zeta_a \psi \phi^\dagger
\label{Portalcoupling}
\eeq
where $N$ is the total number of F-terms, $f_i$, in all sectors, $\tilde{m}^2_i$ is the soft mass contribution from the $i$'th hidden sector F-term, $\tilde{m}^2_i = -f^2_i/{\Lambda_i^2}$, the effective mediation scale of the $i$'th hidden sector to the SSM is $\Lambda_i$,
and $V_{ia}$ is the rotation matrix that diagonalises the goldstini mass matrix.\footnote{To avoid confusion, note that the $f_i$ have
mass-dimension two.}  The $\zeta_a$ are the $N-1$ goldstini mass eigenstates and the true global goldstino that forms the longitudinal component of the gravitino is the $N^{th}$ eigenstate with zero mass in this basis.   
If we make the reasonable assumption that SUSY breaking from all sectors is not communicated in an identical way, \ie, if not {\it all} $\Lambda_i$ are equal, then couplings of the SSM to all goldstini are generated by the interaction of Eq.(\ref{Portalcoupling}). Crucially, these couplings between the goldstini and SSM are parametrically stronger than gravitational, even if the effective mediation scale is $\sim M_P$. Moreover, these goldstini-SSM couplings distinguish the goldstini from other $m_{3/2}$-scale fermions such as, e.g., derivatively-coupled modulini, whose couplings to SSM states at a scale $E$ are suppressed relative to those of goldstini by $E \Lambda_i / f_i$.\footnote{Though this is true of conventional, derivatively coupled modulini, there are of course exceptions -- for example, the fermionic components of moduli superfields involved in supersymmetry breaking, whose couplings to SSM states are goldstino-like. We particularly thank Joe Conlon for discussions on related issues.} 

For the field-theoretic breaking of global SUSY, it is reasonable to expect that the distribution of breaking scales is roughly log-flat since, assuming SUSY is unbroken at tree level, breaking only occurs via non-perturbative
effects (modulo technicalities involving Fayet-Illiopoulos terms), which scan over an exponentially large range of scales as UV couplings and beta-function coefficients are linearly changed
\cite{Witten:1981nf,Intriligator:2007cp}.  In the context of multi-sector SUSY breaking, there should be a lower cutoff on this distribution of SUSY-breaking scales, implied by the (at least gravitational strength, \ie, anomaly-mediated) communication of breaking from the dominant SUSY-breaking sector to the sub-dominant ones.  If we require SUSY to solve the hierarchy problem, and we assume the dominant SUSY breaking  sits at the intermediate scale, the lowest independent SUSY-breaking sector should have scale $\sim \tev$.  In Section \ref{mass} we will see how such a scenario leads to a strong modification of the goldstino mass when arising from a sector with such a very low SUSY-breaking scale.

In fact, in the landscape of string theory, one might naively expect `tree-level' breaking  due to the presence of fluxes or anti-D-branes in the vacuum not to be distributed at all scales, but instead concentrated at the string scale.  Nevertheless, because of the presence of warped throats (caused by the back reaction from fluxes or branes), an approximately log-flat distribution of SUSY-breaking scales can still apply due to the approximately log-flat distribution of throat lengths expected in realistic string compactifications\cite{Giddings:2001yu,Dimopoulos:2001qd,Hebecker:2006bn}, and such structures are further motivated by the phenomenological necessity of conformal sequestering if
SUSY is relevant to the solution of the hierarchy problem.   In this context it is also noteworthy that anti-D-branes (or equivalent fluxes) sitting at the IR tip of one or more
throats can have significant utility in the string landscape, as the presence of the anti-D-brane charges relaxes the tadpole constraints on
the allowed vacua, and thus allows for a (quite possibly exponentially) larger landscape of vacua.   

Having discussed our view of the overall scene in which the goldstini scenario is set and motivated, we now turn to our specific results, starting with the changes to the goldstini mass spectrum related to the stabilization of SUSY breaking vacua. 

\section{Goldstini Masses in Supergravity \label{mass}}

Perhaps the clearest way to study the goldstino mass spectrum in supergravity is through the use of the conformal compensator formalism \cite{Cremmer:1978hn, Cremmer:1982en,Luty:2000ec}. The relevant physics may be captured by considering a single chiral superfield $X(y) = x(y) + \sqrt{2} \psi_X(y) \theta + f_X(y) \theta^2$ with a Polonyi-type superpotential and \kahler terms necessary for stabilizing the vacuum at finite $\langle x \rangle$. The Lagrangian is given by 
\beq
\mathcal{L} =\int d^4\theta \, \phi^\dag \phi \left(  X^\dagger X - \frac{c (X^\dagger X)^2}{M^2} +... \right) + \int d^2 \theta \phi^3 f X + h.c.
\eeq
where $\phi = \phi + f_\phi \theta^2$ is the conformal compensator and $c>0$. Such a Lagrangian naturally arises as an effective description of SUSY breaking valid below the scale $M$ (e.g., an O'Raifeartaigh model with fields of mass $M$)\footnote{Note that here, for simplicity, we have assumed an R-symmetry is preserved.  Inclusion of \kahler terms such as $c' X^\dag X^3/M^2$ allow the study of R-breaking cases, with similar results to the R-preserving case.}. The quartic stabilizing term in the \kahler potential is absolutely necessary in the context of supergravity; its absence would induce a runaway to large field values. If $X$ were the only source of supersymmetry breaking, we would identify $\psi_X$ as the true longitudinal goldstino $G$ that is eaten by the gravitino. Indeed, in this case the zero momentum equation of motion for $x$ may be solved to yield $x = \frac{\psi_X^2}{2 f_X}$. Thus in the far infrared we may write $X$ as a nonlinear superfield,
\beq
X = \frac{G^2}{2 f_X} + \sqrt{2} G \theta + f_X \theta^2
\eeq
which corresponds to the usual nonlinear parameterization of the goldstino $G \equiv \psi_X$ \cite{Komargodski:2009rz}. 

Now let us consider the effects of multiple SUSY breaking on the fermion $\psi_X$. We assume the dominant contribution to SUSY breaking comes not from $X$, but from other sectors sequestered from $X$, so that $\langle f_\phi / \phi \rangle = m_{3/2}$. Clearly, it is now necessary to keep careful track of dependence on the conformal compensator. We may analyze the effects of SUSY breaking on $X$ by going to the canonical basis via the rescaling $X \to X / \phi$ and solving the auxiliary equation of motion to find
\beq
f_X = - \frac{2 c (f_\phi/\phi) |x|^2 x + 2 c x^\dag \psi_X^2 + f M^2}{M^2  - 4 c |x|^2}.
\eeq
By minimizing the resulting scalar potential for $x$, we may then extract the mass for the would-be goldstino $\eta \equiv \psi_X$,
\beq \label{realmass}
m_\eta = 2 m_{3/2} \left ( 1 - \frac{M^2 m_{3/2}^2}{2 c f^2} + ...  \right)
\eeq
where additional correction terms are $\mathcal{O} \left( \frac{M^4 m_{3/2}^4}{c^2 f^4} \right)$. This expansion is valid in the regime $m_{3/2}^2 / c \ll f^2/M^2 \lsim m_{3/2} M_P.$ 
One can see that as $\sqrt{f}$ (and $M$) approaches $m_{3/2}$ these corrections become significant and a different expansion is necessary.  From a numerical study we find that for $\sqrt{f} < m_{3/2}$ these large corrections can drive the goldstini mass much smaller than $2 m_{3/2}$.  Such corrections are to be expected as in this case the SUSY-breaking communicated to a sector becomes larger than the breaking within the sector itself and there is no SUSY to be spontaneously broken within the sector from the outset.

A few remarks are in order. As can be seen clearly in Eq.(\ref{realmass}), the goldstino mass in a given SUSY-breaking sector depends on both the overall scale of SUSY breaking and the scales within the sector itself; the interplay of supersymmetry breaking, gravitational effects and vacuum stabilization leads to important corrections to the goldstino mass - whenever the SUSY breaking field $X$ gains a large vacuum expectation value linear terms in the \kahler potential lead to mixing between $X$ and the gravity multiplet, and thus corrections to the goldstino mass\footnote{We thank Clifford Cheung and Jesse Thaler for discussions concerning this interpretation and for correcting a numerical factor in the original version of Eq.(\ref{realmass}).  Corrections due to mixing with the gravity multiplet are also discussed in the Appendix A of \cite{Goldstini1}.}.  Of course, the generalization of this setup to $N$ SUSY-breaking sectors is straightforward, resulting in $N$ goldstini $\eta_i$; in the mass eigenbasis these become the eaten longitudinal goldstino and $N-1$ uneaten  goldstini $\zeta_a$ (related to the $\eta_i$ by $\eta_i = V_{ia} \zeta_a$, where $V_{ia}$ is the rotation matrix that diagonalizes the goldstini mass matrix).

In this case it may seem that if one makes a unitary transformation such that there is only one Polonyi field $G = \sum_i f_i X_i / f_{eff}$ all other orthogonal combinations $\tilde{X}_i$ might remain massless by this derivation.  However, in this new basis the stabilizing K\"{a}hler term will lead to mixed interactions between $G$ and the other fields $\tilde{X}_i$.  The non-zero vev of $G$ then leads to masses for the fermionic components of $\tilde{X}_i$ and the same results are recovered.

In subsequent sections, we will often be interested in computing corrections to $m_\eta = 2 m_{3/2}$ due to additional physics such as warping and sequestering. In such cases, for convenience we will dispense with the details of stabilization and instead carry out a na\"{i}ve application of the nonlinear goldstino parameterization 
\beq
X = \frac{\eta^2}{2 f_X} + \sqrt{2} \eta \theta + f_X \theta^2
\eeq
 for the would-be goldstino $\eta$ given a Polonyi superpotential and canonical \kahler term for $X$. For free fields without warping this parameterization gives the leading result $m_\eta = 2 m_{3/2}$, which omits the corrections due to mixing with the gravity multiplet but is valid in the limit $m_{3/2}^2 / c \ll f^2/M^2$. Such a simplifying assumption will make the effects of new physics more transparent, with the understanding that corrections from stabilization have been suppressed.
 
Let us now turn to one such genre of new physics -- the changes to the goldstini mass spectrum arising from the warping and/or conformal sequestering that can naturally occur in the string landscape.

\section{Warped and Sequestered Goldstini}\label{warpandseq}

The observed smallness of FCNCs require that SUSY breaking communicated via flavour-violating mechanisms such as gravity mediation must necessarily be subdominant to flavor-preserving contributions. Absent some degree of unnatural tuning, this is most readily achieved by warping on an extra-dimensional space \cite{Randall:1998uk}  or, in four dimensions (and essentially equivalent by AdS/CFT duality) sequestering by a conformal sector \cite{Luty:2001jh, Luty:2001zv}. Such sequestering is known to arise readily in string theory in the presence of strongly warped backgrounds \cite{Kachru:2007xp}. But even apart from considerations of flavour, the persistence of multiple goldstini requires that different SUSY breaking sectors be sequestered from each other in a similar fashion; in fact, one should think of sequestering and multiple goldstini as inextricably intertwined. Given the effective dimensional transmutation brought about by warping, it is then natural to consider whether the scale of goldstino masses may be significantly modified if the additional SUSY breaking sector is at the bottom of a warped throat or in the far IR of a pseudo-conformal sector.

As before, the conformal compensator formalism may be used to clearly study the effects of warping or sequestering on the goldstino mass prediction, $m_\eta = 2 m_{3/2}$.  To get started, consider some number of chiral superfields $X_i$ with Polonyi-type superpotentials and a sequestered K\"{a}hler potential. The relevant Lagrangian is 
\beq
\mathcal{L} = \int d^4 \theta \; \phi^\dag \phi \sum_i (X_i^\dag X_i + ...) + \int d^2 \theta \; \phi^3 \sum_i \mu_i^2 X_i + \text{h.c.}
\eeq
where, for simplicity, we have omitted terms necessary for stabilizing the SUSY breaking vacuum.\footnote{As mentioned above, this omits the leading corrections to $m_\eta$ related to stabilization found in Eq.(\ref{realmass}).}  Here $\phi = \phi + f_\phi \theta^2 $ is the appropriate conformal compensator, which, as we will see below, need {\it not} always be identified with the SUGRA conformal compensator. Rescaling $X_i \to X_i / \phi$ and expanding $X_i$ in the nonlinear parameterization
\beq
X_i =  \eta_i^2/2 f_{X_i} + \sqrt{2}  \eta_i \theta + f_{X_i} \theta^2
\eeq
we obtain
\beq
\mathcal{L} \supset \int d^2 \theta \; \phi^2 \sum_i \mu_i^2 X_i = - \frac{1}{2} \left( 2 \frac{f_\phi}{\phi} \right) \sum_i \eta_i^2 + \text{constant}
~~. 
\label{eq:nonlineargold}
\eeq

There are two salient details worth noting in this result. The first is that here we have assumed the $X_i$ are free fields with canonical scaling dimension; as we will discuss  below, the result changes significantly when the scaling dimensions of fields responsible for SUSY breaking differ from unity.  The second is that the conformal compensator is ultimately responsible for setting the goldstino masses.  The additional Goldstini of multiple SUSY breaking obtain masses of order $m_\eta = 2 \frac{f_\phi}{\phi}$; this only corresponds to $m_\eta = 2 m_{3/2}$ when $f_\phi / \phi = m_{3/2}$, which is not guaranteed to be the case, as we will shortly show.

Perhaps the simplest example of such deviations arise when the some of the chiral fields $X_i$ possess  scaling dimensions $\Delta_{X_i} > 1$, possibly at a conformal or near-conformal fixed point.\footnote{In order to compare results with those in \cite{Goldstini1}, we will use the conventions $\Delta_X = 1 + \gamma_X$ to define the anomalous dimension $\gamma_X$ in terms of the scaling dimension $\Delta_X$. This corresponds to the choice $d \ln Z_X / d \ln \mu_R = - 2 \gamma_X$.} Such circumstances arise frequently in theories with conformal sequestering \cite{Luty:2001jh, Luty:2001zv, Schmaltz:2006qs, Kachru:2007xp}, and more generally whenever SUSY breaking sectors are strongly coupled.

To see the effects of large anomalous dimensions more clearly, let us focus on the case of a single chiral superfield $X$ with scaling dimension $\Delta_X \neq 1$ at a conformal fixed point. It is frequently the case that $X$ is a component of a gauge invariant
chiral operator of some interacting gauge theory, e.g., an $SU(2)$ theory with moduli space of gauge-invariant
meson operators parametrized as
\beq
M = \left( \begin{array}{cc} \epsilon Z & X \\
-X^T & \mathcal{O}(X^2/Z) \end{array} \right)~~.
\eeq
Expanding around $Z\neq 0, X = 0$, conformal symmetry demands that the K\"{a}hler potential for $X$ must be of the form 
\beq
K = \phi^\dag \phi (Z^\dag Z)^{1/\Delta_Z}\left[1 + \frac{ X^\dag X}{(Z^\dag Z)^{(\Delta_X/\Delta_Z)}}  + ... \right]~~,
\eeq
where $\phi$ is the SUGRA conformal compensator with $\vev{\phi} = 1 + m_{3/2} \theta^2$.  We assume there is also a
superpotential Polonyi term
\beq
W = \phi^3 \mu^2 X
\eeq
where the constant $\mu^2$ has dimension $(3-\Delta_X)$; additional K\"{a}hler terms are required, as usual, to stabilize the SUSY breaking vacuum of $X$.  We can study the theory near the origin of moduli space in terms of redefined fields 
\beq
\hat Z = \phi Z^{1/\Delta_Z}\quad {\rm and}\quad \hat X = \phi^{\Delta_X} X/\hat Z^{\Delta_X - 1}~~,
\eeq
for which the \kahler potential is canonical (without any dependence on $\phi$) and the superpotential term becomes
\beq
W \to \phi^{2-\gamma_X} \mu^2 \hat Z^{\gamma_X} \hat X~~.
\eeq
We are interested in the mass term for the goldstino component of $\hat X$. Utilizing the nonlinear parameterization
of Eq.(\ref{eq:nonlineargold}), we find, in the case of current interest,
\beq
\lang \supset - \half (2-\gamma_X) m_{3/2} \eta^2~~,
\eeq
from which we see the goldstino mass is
\beq
m_\eta = (2 - \gamma_X) m_{3/2}~~,
\label{eq:modifiedGmass}
\eeq
in agreement with the perturbative result of \cite{Goldstini1}.  The key point here is that $\gamma_X$ need not be perturbative, so in principle the goldstino mass may range from $0$ to $2 m_{3/2}$ depending on the size of $\gamma_X$. For example, if the superpotential of the gauge theory at the interacting fixed point involved a marginal operator, $W \supset \text{tr} M^2$, we would have $\gamma_X = 1/2$ and thus $m_\eta = \frac{3}{2} m_{3/2}$.

This is a simple example of our first point -- that the mass of a goldstino coming from a chiral superfield $X$ depends sensitively on the scaling dimension $\Delta_X$. The smallness of FCNCs suggests that multiple SUSY-breaking sectors, if present, must be sequestered in order to avoid prohibitive flavour-violating contributions to soft masses. In four dimensions, this is most readily accomplished by conformal sequestering, in which case anomalous dimensions $\gamma_X \neq 0$ are generically expected.

Thus far our discussion has also assumed that the mediation of dominant SUSY breaking arises through the conventional SUGRA conformal compensator; as we will now argue, this, too, no longer holds in many situations where SUSY breaking fields are sequestered by conformal dynamics or warping in higher-dimensional spaces.

\subsection{Warped goldstini 5D}

In order to probe the effects of warping on goldstino masses, let us consider a toy model of warping in the form of a 5D supersymmetric Randall-Sundrum model \cite{Randall:1999ee}. While such constructions are perhaps not as realistic as those based on more complete warped throat solutions \cite{Klebanov:2000nc,  Klebanov:2000hb}, they nonetheless capture much of the relevant physics.  To set notation, we take the 5th dimension  to be compactified on an interval of length $\pi r$ via a $S^1/Z_2$ orbifold, with metric
\bea
ds^2 = e^{-2kr |\theta|} \eta_{\mu \nu} dx^\mu dx^\nu + r^2 d \theta^2
\eea
for $-\pi < \theta \leq +\pi$; the slope discontinuities at $\theta = 0, \pi$ signal the presence of 4D branes fixed by the orbifold boundary conditions.  These branes mock up the resolved physics of the UV Calabi-Yau `head', and IR throat `tip' in the IR of the more realistic
complete string solutions.  As usual, the warp factor $e^{-2 k r |\theta|}$ indicates that physical scales on the $\theta = \pi$ IR brane are redshifted relative to those on the $\theta = 0$ UV brane.

At energies below the mass of the lightest gravitational Kaluza-Klein (KK) mode, we may employ an effective 4D Lagrangian describing the physics of fields localized on UV and IR branes separated by a warped throat. The Lagrangian for this effective theory is \cite{Luty:2000ec, Bagger:2000eh} 
\bea
\lang = - \frac{3 M_5^3}{k} \int d^4 \theta \left( \phi^\dag \phi - \omega^\dag \omega \right) + \int d^4 \theta (\phi^\dag \phi K_{UV} + \omega^\dag \omega K_{IR} ) \\ \nonumber
+\int d^2 \theta (\phi^3 W_{UV} + \omega^3 W_{IR}) + h.c.
\eea
Here $\phi$ is the conformal compensator field and $\omega$ is the ``warp factor'' superfield,
\beq
\omega = \phi e^{-kT}
\eeq
where $T = \pi r + ...$ is the radion superfield (in a horrible abuse of notation, we will write the warp factor superfield in terms of its scalar and auxiliary components as $\omega = \omega + f_\omega \theta^2$). The physics we are interested in will be encoded by \kahler and superpotential terms for a Polonyi field localized on the IR brane. The anomaly-mediated communication of supersymmetry breaking to fields localized in the IR arises via the warped conformal compensator $\omega$, giving rise to supersymmetry breaking of order $f_\omega/ \omega$. 

Ultimately, the size of supersymmetry breaking seen by IR fields is determined by the stabilization mechanism fixing the expectation value of the radion, and hence the warp factor superfield $\omega$. Although it is often the case that simple forms of radius stabilization lead to $\langle f_\omega / \omega \rangle \sim \langle f_\phi \rangle$ (as in, e.g., \cite{Luty:2000ec}), we will be interested in a much more general class of stabilization mechanisms.
 
Now let us consider the effects of this stabilization on supersymmetry breaking in the IR. Suppose that the field content in the IR includes one or more fields breaking supersymmetry (in addition to other sources of supersymmetry breaking in the UV). We may represent this locally by a Polonyi model for a field $X$, via a superpotential term $W_{IR} = \mu^2 X + ...$ and  K\"{a}hler term $K_{IR} = X^\dag X$ (along with the usual additional K\"{a}hler terms necessary to stabilize the potential). Assume now that the dominant contribution to supersymmetry breaking arises elsewhere on the manifold, so that $X$ can be identified as a non-linear pseudo-goldstino field. To study the dynamics of $X$, we may rescale $X \to X / \omega$, which results in canonical K\"{a}hler terms for $X$ and a superpotential
\beq
W_X = \omega^2 \mu^2 X ~~.
\eeq
The resulting goldstino mass term is
\beq
\lang \supset \omega f_\omega \frac{\mu^2}{f_X} \eta^2 = - \half (2 f_\omega / \omega) \eta^2~~.
\eeq
As expected, the mass for this IR-localized goldstino depends on the warped SUSY-breaking order parameter $\langle f_\omega / \omega \rangle$ rather than the UV order parameter $\langle f_\phi \rangle$.

What are the effects of warping on the goldstino mass spectrum?  Clearly, in the case of no warping, $\omega = \phi = 1 + m_{3/2} \theta^2$ and hence $m_\eta = 2  f_\omega/\omega = 2 m_{3/2}$, consistent with the familiar result. Moreover, in the event that there is nontrivial warping but the stabilization mechanism yields $f_\omega / \omega \sim f_\phi = m_{3/2}$, we again obtain $m_\eta \simeq 2 m_{3/2}$. However, this is far from the only possible outcome.  Consider a stabilization superpotential of the form \cite{Luty:2002ff}
\beq
\lang = \int d^2 \theta (c_{UV} \phi^3 + c_{IR} \omega^3 + \epsilon \phi^{3-n} \omega^n) + h.c - f_{UV}^2 [1+\text{gravity terms}]~~.
\eeq
The first two terms can arise from constant superpotentials localized on the UV and IR branes; the third term requires a bulk gauge theory with some massive fundamental matter. 

For $\epsilon \ll c_{UV}, c_{IR}$ and $n < 3$, the $\epsilon$ term contributes a vev to $\omega$ of order 
\beq
|\langle \omega \rangle |^{4-n} = \frac{n(3-n)}{6} \left| \frac{\epsilon c_{UV}}{c_{IR}^2} \right| \ll 1~~,
\eeq
with 
\bea
\left | \frac{\langle f_\omega \rangle}{\vev{\omega}} \right| = \frac{|c_{IR}|}{M_P^2} |\vev{\omega}| \quad {\rm and} \quad |\vev{f_\phi}| = \frac{|c_{UV}|}{M_P^2} = \frac{f_{UV}}{\sqrt{3} M_P}  ~~.
\eea
Here the radion mass is of order $\langle f_\omega / \omega \rangle$, while the gravitino mass is of order $\langle f_\phi \rangle$. Significantly, the order parameter for SUSY breaking in the IR is parametrically suppressed relative to $\langle f_\phi \rangle$. 
Thus, in this case, the goldstino mass is
\beq
m_\eta = 2 \frac{f_\omega}{\omega} = 2 \frac{|c_{IR}|}{M_P^2} |\vev{\omega}| \ll 2 m_{3/2}~~.
\eeq
Depending on the choice of stabilization parameters, this results in a goldstino mass ranging between $0 < m_\eta \leq 2 m_{3/2}$.  The generalization to many Goldstini is straightforward; for $\eta \to \eta_i$ one need only take $\omega \to \omega_i$ and $f_\omega \to f_{\omega_i}$ in the case of multiple throats.

\subsection{Sequestered goldstini in 4D}

As one might expect, we can also see the effects of warping on goldstino masses in a strictly four-dimensional picture of conformal sequestering. In this situation the role of warping is played by the dynamics of a superconformal sector coupling to the IR fields.
Following \cite{Luty:2002ff}, for the sake of specificity we will focus on the case of a 4D $SU(2)$ SUSY gauge theory with 8 fundamentals $P$ and superpotential
\beq
W = \lambda P^4 + \kappa P^2.
\eeq
This theory flows to a conformal fixed point in the infrared, where the coupling $\lambda$ is assumed to be marginal. At the conformal fixed point, the superconformal $R$-symmetry fixes the scaling dimension of $P$ such that $\Delta_P = 3/4$; thus $\lambda$ is dimensionless (marginal) and the coefficient $\kappa$ has scaling dimension $\Delta_\kappa = 3/2$. The moduli space of gauge invariant operators can be parameterized as
\beq
PP = \left( \begin{array}{cc} \epsilon Z & Y \\
-Y^T & \mathcal{O}(Y^2/Z) \end{array} \right)
\eeq
where $\epsilon = i \sigma_2$ is the antisymmetric tensor. Here $Y$ is a $2 \times 6$ matrix of fields. 

Conformal symmetry constrains the theory below the scale $Z$ to have K\"{a}hler terms of the form
\bea
K = \phi^\dag \phi (Z^\dag Z)^{2/3} [1+ \mathcal{O}(|Y|^2/|Z|^2)] \\ \nonumber
= \hat Z^\dag \hat Z [1+\mathcal{O}(|\hat Y|^2/|\hat Z|^2)] ]
\eea
where $\hat Z = \phi Z^{2/3}$ and $\hat Y = \phi Y / Z^{1/3}$. In terms of these variables, the superpotential becomes 
\beq
W = \lambda \hat Z^3 + \kappa \hat Z^{3/2} \phi^{3/2}~~,
\eeq
which has the same form as our 5D Randall-Sundrum theory with $n = 3/2$ and $\hat Z \sim \omega$. Indeed, if we make the identifications $\hat Z \to M_P \omega$, $\lambda \to c_{IR} / M_P^3$, and $\kappa \to \epsilon / M_P^{3/2}$, we may reproduce all the details of the warped model in terms of a four dimensional conformal field theory.

Of course, we may consider a wide range of conformal field theories with various marginal operators at  the conformal fixed point. In general, a superpotential
\beq
W = \lambda P^k + \kappa P^2
\eeq
leads, below the scale of $Z$, to an effective superpotential
\beq
W = \lambda \hat Z^3 + \kappa \hat Z^{6/k} \phi^{3 - 6/k}
\eeq
where $n = 6/k$. The constraint $n > 3$ corresponds to $k > 2$; for $k > 2$, $\Delta_Z< 3$, which is eminently sensible in order that $\kappa$ remain a relevant deformation. In any event, we need not commit to a specific conformal field theory; any dynamics with $\Delta_Z > 1$ may suffice.

Now let us consider the effects of sequestering on Goldstini coupled to the conformal sector. This corresponds to coupling the field $\hat Z$ (which is our stand-in for the warp superfield) to a field with a Polonyi term. First, consider the theory where the field $X$ that breaks SUSY is a total composite of scaling dimension $\Delta_X = 3$. This is the four dimensional analog of a purely ``IR-localized'' field; the case of $\Delta_X < 3$ corresponds to a partially-localized field, which we will discuss momentarily.

Expanding around $Z \neq 0$ and $X = 0$, the \kahler potential below the scale $Z$ is constrained by conformal symmetry to be of the form
\beq
 K \supset \phi^\dag \phi (Z^\dag Z)^{1/\Delta_Z} \left[1 + \frac{X^\dag X}{(Z^\dag Z)^{(3/\Delta_Z)}} + ... \right]
\eeq
We may thus define canonical fields 
\beq
\hat Z = \phi Z^{1/\Delta_Z} \quad {\rm and} \quad \hat X = \phi^3 X / \hat Z^2
\eeq
in terms of which the \kahler potential is canonical.

If the theory contains a Polonyi term for the candidate SUSY breaking field $X$, the superpotential is of the form
\beq
W = \phi^3 \mu^2 X \to \mu^2 \hat Z^2 \hat X
\eeq
It is then a simple matter to compute the goldstino mass; the Lagrangian includes a term
\beq
\lang \supset - \half \left(2 \frac{f_Z}{\hat z} \right) \eta^2
\eeq
so that the goldstino mass is given by $m_\eta = 2 f_Z / \hat z \sim 2 f_\omega / \omega$, as expected from the results of the previous subsection. The stabilization mechanism for $Z$ is simply the one considered earlier.

We may also consider the case where $1 \leq \Delta_X \leq 3$, i.e., the candidate SUSY breaking field has a large anomalous dimension but should not be interpreted as being completely localized on the brane; rather, it has a warped profile in the 5D picture corresponding to a bulk mass term.

Once again, expanding around $Z \neq 0$ and $X = 0$, the \kahler potential is constrained by conformal symmetry to take the form
\beq
K \supset \phi^\dag \phi (Z^\dag Z)^{1/\Delta_Z} \left[1 + \frac{X^\dag X}{(Z^\dag Z)^{(\Delta_X/\Delta_Z)}} + ... \right]~~,
\eeq
with canonical fields given by $\hat Z = \phi Z^{1/\Delta_Z}$ and $X = \phi^{\Delta_X} X / \hat Z^{\Delta_X - 1}$,
in terms of which the \kahler potential is canonical.   The superpotential term for $X$ thus takes the form
\beq
W = \phi^3 \mu^2 X \to  \mu^2 \phi^{2- \gamma_X} \hat Z^{\gamma_X} \hat X~~.
\eeq
Carrying out the nonlinear parameterization for the goldstino, we find in this case a goldstino mass
\beq
m_\eta = (2 - \gamma_X) m_{3/2} + \gamma_X \frac{f_Z}{\hat z} ~~.
\label{eq:generalGmass}
\eeq
This result interpolates nicely between the results found in the limiting cases $\gamma_X = 0$ and $\gamma_X = 2$.  In the former limiting case, we retrieve the physics of a free Polonyi field with no warping; in the latter limiting case, the physics of a fully sequestered Polonyi field where the scale of SUSY breaking is set not by $f_\phi$, but $f_\omega / \omega$. 

Thus far we have remained relatively agnostic about the detailed physics of supersymmetry breaking, but this, too, may have a significant impact on the spectrum of goldstini, as we will now see.

\section{Multiple Goldstini and Modulini from ISS Sectors}\label{ISS}

The notion of multiple SUSY breaking sectors prompts us to consider how SUSY may be broken within each sector.  The ISS models \cite{ISS} demonstrate that SQCD with massive flavours exhibits a meta-stable SUSY breaking ground state.  Further, the simplicity of such models would suggest that spontaneously broken SUSY is generic in SUSY field theory and in the landscape of string vacua.  Therefore it is natural to consider, in the context of multiple SUSY breaking sectors, that some number may well be of the ISS type, without the addition of any of the singlets or deformations that are absent in the original ISS models, and that are needed only to break $R$-symmetries.  Here we show that such a sector would give rise to multiple goldstini fields along with many more `modulini' fields of mass $\geq 2 m_{3/2}$.\footnote{Purely for typographical clarity we ignore, throughout this section, the possibility of the warping or conformal sequestering considered in Section \ref{warpandseq}.  We emphasise that typically the metastable
SUSY-breaking of ISS-type studied in the present section should also come along with such sequestering dynamics, leading to the changes
in overall goldstini (and modulini) mass scales and couplings explicated in Section \ref{warpandseq}.}     These extra states could potentially lead to a smoking gun signature of an ISS hidden sector by determining missing energy in LOSP decays to the gravitino, goldstini and modulini.

\subsection{ISS models at low energies}\label{ISSatlowE}

To illustrate the essential physics we concentrate on the classic ISS-model of SQCD with $N_c$ colours and $N_c + 1 \leq N_f < \frac{3}{2} N_c $ flavours in the free magnetic range \cite{Seiberg:1994bz,Seiberg:1994pq,Intriligator:1995au}.  The generalization to other gauge groups should be straightforward.   A simple, intuitive understanding of why such an ISS sector gives rise to multiple goldstini fields comes from the fact that, in the far IR, it flows to multiple decoupled O'Raifeartaigh-like models as we now show.

Using Seiberg duality \cite{Seiberg:1994pq} the IR-free description of the theory is described by an $N_f \times N_f$ gauge singlet meson matrix $\Pi_{ij}$ and $N_f$ flavours of magnetic quarks $\bold{\varphi}_i$ and $\bold{\tilde{\varphi}}_j$ in the fundamental (respectively anti-fundamental) of a $SU(\tilde{N} = N_f-N_c)$ magnetic gauge theory.
This theory is weakly coupled at low energies and has a superpotential given by
\begin{equation}
W = h \Tr\left[\bold{\varphi} \cdot \Pi \cdot \bold{\tilde{\varphi}} - \mu^2 \cdot \Pi \right].
\end{equation}
We assume a generic, non-hierarchical, matrix $\mu^2_{ij}$ which can be diagonalized without loss of generality.  Among other symmetries this theory exhibits a $U(1)_R$ symmetry where the $\varphi$ fields have zero R-charge and $\Pi$ has R-charge 2.

Considering the F-components of the meson superfields,
\begin{equation}
-F^\dag_{\Pi_{ij}} = h \bold{\varphi}_i \cdot \bold{\tilde{\varphi}}_j - h \mu^2_{ij}
\end{equation}
the first term in this matrix equation is of rank $N_f-N_c$ whereas the second term is of rank $N_f > N_f-N_c$, therefore it is impossible to have $F_{\Pi_{ij}} = 0$ for all $\{i,j\}$ and SUSY is broken.  This is the famous ISS `rank condition'.  The minimum of the potential is
\begin{equation}
V = \sum_i^{N_c} (h \mu^2_i)^2
\end{equation}
where $\mu^2_i$ are the $N_c$ smallest eigenvalues of $\mu^2_{ij}$.  This minimum occurs in field space
\begin{equation}
\Pi = \left( \begin{array}{cc}
Y & Z \\
\tilde{Z} & \Phi \end{array} \right),  \quad 
\varphi = \left( \begin{array}{cc}
\varphi_0+  \chi, & \rho \end{array} \right), \quad 
\tilde{\varphi} = \left( \begin{array}{c}
\tilde{\varphi}_0 +  \tilde{\chi} \\
 \tilde{\rho} \end{array} \right), \quad 
\mu^2 = \left( \begin{array}{cc}
\tilde{\mu}^2_0 & 0 \\
0 & \mu^2_0 \end{array} \right)
\end{equation}
with $\bold{\varphi_0} \cdot \bold{\tilde{\varphi_0}} = \tilde{\mu}^2_0$.  Also, $\Phi$ is an $N_c\times N_c$ matrix of fields, $Y$ is $(N_f-N_c)\times (N_f-N_c)$, $\rho$ is $N_c\times N_c$ and the dimensionality of the other terms is apparent from these assignments.  Upon rewriting the superpotential in terms of these fields it splits
into three pieces $W=W_1 + W_2 + W_3$ with
\bea
W_1 & = &  h \Tr[\rho \cdot \Phi \cdot \tilde{\rho} + \rho \cdot \tilde{Z} \cdot \tilde{\varphi}_0 + \varphi_0 \cdot Z \cdot  \tilde{\rho} - \mu_0^2  \cdot \Phi] \nonumber \\
 & = & - h \Tr[\mu_0^2  \cdot \Phi] + h \sum_{i=1}^{N_f-N_c} (\boldsymbol{\phi}_{1_i} \cdot \Phi \cdot \boldsymbol{\phi}_{2_i} + \tilde{\mu}_{0_i} (\boldsymbol{\phi}_{1_i} \cdot \boldsymbol{\phi}_{4_i}+\boldsymbol{\phi}_{2_i} \cdot \boldsymbol{\phi}_{3_i})) ~~.
 \label{SuPlow}
\eea
Here the $\boldsymbol{\phi}$ are $N_c$ dimensional vectors, and the $\tilde{\mu}_{0_i}$ are the first $N_f-N_c$ diagonal components of the $\tilde{\mu}_0^2$ matrix. 
In the first line we recognise $W_1$ as an O'Raifeartaigh-like model and in the second line the fields $\rho, \tilde{\rho}, \tilde{Z}$, and $Z$ have been written as matrices made up of row and column vectors to demonstrate explicitly how the superpotential $W_1$ decomposes into $N_f-N_c$ O'Raifeartaigh-like sectors.   The remaining pieces of the superpotential are
\bea
W_2 & = & h \Tr[ \chi \cdot Y \cdot  \tilde{\chi}  + \chi \cdot Y \cdot \tilde{\varphi}_0 + \varphi_0 \cdot Y \cdot \tilde{\chi}] \\
W_3 & = &h \Tr[ \rho \cdot  \tilde{Z} \cdot \tilde{\chi}  +  \chi \cdot Z \cdot \tilde{\rho}] ~~.
\eea
$W_2$ comprises a sector which doesn't break SUSY and contains massive chiral superfields along with the Goldstone superfields of the spontaneously broken symmetries.  The Goldstone fields of the spontaneously broken $SU(N_f-N_c)$ are eaten by the gauge superfields through the supersymmetric Higgs mechanism.  These SUSY-preserving fields are only coupled to the SUSY-breaking sector through the cubic terms in $W_3$ and can therefore be consistently neglected when considering the first sector.

It is clear that $\Phi$ remains massless at tree level, and the diagonal component of $\Phi$ contains the goldstino.  The pseudo-moduli of this field become massive at one-loop level through their interactions with the heavy $\boldsymbol{\phi}$ fields and these masses can be calculated to all orders in the SUSY breaking parameters with the use of the Coleman-Weinberg potential \cite{Coleman:1973jx}.  However, as we would later like to embed this theory in SUGRA, and the Coleman-Weinberg approach is not manifestly supersymmetric, we choose instead to work in terms of the effective K\"ahler potential which arises when the heavy superfields are integrated out.  This agrees with the Coleman-Weinberg potential to second order in the SUSY breaking F-terms and in the limit where SUSY is unbroken this is exact at one-loop.\footnote{Including higher order corrections in the SUSY breaking parameters would necessitate including supercovariant derivates.  The effective K\"ahler potential is sufficient for our needs.}

In general for a superpotential of the form
\beq
W = \frac{1}{2} M_{ij} \bold{\varphi}_i \cdot \bold{\tilde{\varphi}}_j~~,
\label{Mdef}
\eeq
where $M_{ij}$ includes mass terms and the pseudomoduli fields, the exact one-loop K\"ahler potential is given by
\beq
K^{(1)} = - \frac{1}{32 \pi^2} \Tr\left[M^\dagger M \log\left(\frac{M^\dagger M}{|\Lambda|^2}\right)\right]~~.
\label{keff}
\eeq

Reading off the matrix $M$ from Eq.(\ref{SuPlow}) one finds that the theory describing the light fields contained in $\Phi$, after integrating out the heavy fields contained in $\bf{\phi}$, is described by the superpotential
\beq
W  =  h \Tr[\mu_0^2  \cdot \Phi] ~~,
\label{lowtheoryW}
\eeq
and the effective K\"ahler potential $K_{eff}= K^{(0)} + K^{(1)}$, where $K^{(0)}$ is the canonical K\"ahler potential, and $K^{(1)}$ is given by
\beq
K^{(1)} =  - \frac{h^2}{32 \pi^2} \sum_{i=1}^{N_f-N_c} \Tr\left[2\left(2 + \log\left(\frac{|\tilde{\mu}_{0_i}|^2}{\Lambda^2}\right)\right) \Phi^\dagger \cdot \Phi + \frac{1}{3 |\tilde{\mu}_{0_i}|^2} (\Phi^\dagger \cdot \Phi)^2 + ...\right] ~~.
\label{lowtheoryK}
\eeq
Here the ellipses denote higher order terms which we can ignore as we are studying the theory near the origin of field space, $\langle \Phi \rangle \ll\tilde{\mu}_0$.  (The first logarithmic correction to the terms quadratic in $\Phi$ corresponds to one-loop wavefunction renormalization of the fields.)   Eqs.(\ref{lowtheoryW}) and (\ref{lowtheoryK}) are sufficient for studying the low-energy phenomenology of the ISS model.  One can see from the quartic term in the K\"ahler potential that, once the diagonal components of $\Phi$ develop F-terms, a scalar potential for all pseudo-moduli in $\Phi$ is generated, and in these (global) SUSY ISS models all scalars are stabilized at the origin $\langle \Phi \rangle =0$.  

Most importantly for our purposes, this low energy theory respects the R-symmetry detailed earlier, forbidding the fermions in $\Phi$, hereafter called `modulini', from gaining mass. This can also be understood by considering the ISS model before integrating out the massive fields: As there are more fermions with R-charge $Q_R=1$ than with $Q_R=-1$, then, if the vacuum is R-symmetry preserving, not all fermions can obtain a Dirac mass, implying some remain massless.   

One may worry that sub-leading corrections spoil this result.  There exist corrections to the K\"ahler potential of the form $\delta K \sim \Tr[\Phi^\dagger \cdot \Phi]^2 / |\Lambda|^2$ where $\Lambda$ is the strong coupling scale of the theory.  These corrections have interesting consequences when the theory is embedded in SUGRA, however as they respect the R-symmetry, we conclude that, in the global limit, they do not contribute to the modulini masses.  There is also a non-perturbative explicit R-symmety-breaking superpotential term 
\begin{equation}
W = N_c (h^{N_f} \Lambda^{-(3 N_c-2 N_f)} \det[\Pi])^{1/(N_f-N_c)}
\end{equation}
generated by gaugino condensation \cite{Davis:1983mz,Affleck:1983mk,Intriligator:1995au,Argyres:2005cb}.   However it preserves a discrete R-symmetry subgroup larger than $\mathbb{Z}_2$, and thus the modulini remain protected from gaining a mass.  We will return, in the next section, to a discussion of these operators in the context of SUGRA .

In summary, one sees that in the global SUSY limit the metastable supersymmetry-breaking vacuum of SQCD with $N_c$ colours and $N_c + 1 \leq N_f < \frac{3}{2} N_c$ massive flavours flavours contains $N_c^2$ massless modulini (of which $N_c$ are goldstini).  In the next section we show that in local supersymmetry these modulini acquire a mass $\geq 2 m_{3/2}$ (ignoring warping and/or conformal-sequestering). 

\subsection{ISS modulini masses in Supergravity}

In SUGRA with spontaneously broken SUSY, one requires a constant term in the superpotential to cancel the cosmological constant from the non-zero scalar potential.  This constant breaks any continuous R-symmetry and we would expect this to manifest itself in a hidden ISS sector by displacing the minimum of the scalar potential for the pseudo-moduli from the origin, and in turn generating masses for the modulini.

First, by considering Eq.(\ref{lowtheoryK}) we see that when $\vev{\Phi} \neq 0$ modulini masses are indeed generated.  As long as $\vev{\Phi} \ll \tilde{\mu}_0$ the dominant contribution comes from the quartic operator, higher order terms leading to subdominant corrections suppressed by higher powers of $\vev{\Phi}/\tilde{\mu}_0$.  For $f_a \sim \mu^2 \ll M_{P}^2$ for all $a$, we show in Appendix \ref{SUGRAmass} that the
condition $\vev{\Phi} \ll \tilde{\mu}_0$ is satisfied.  Thus our SUGRA analysis is valid whenever the scale of supersymmetry breaking is
parametrically below the Planck mass.

Moreover, as also detailed in Appendix \ref{SUGRAmass}, we find, for fields, $X_i$, with super- and K\"ahler-potentials of the form,
\bea
W & = & W_0 + f_a X_a\\
K & = & X_a X^{\dagger}_{a^\star} + \frac{1}{\mu^2} A_{a b^\star c d^\star} X_a X^{\dagger}_{b^\star} X_c X^{\dagger}_{d^\star}~~,
\label{KandW}
\eea
and under the same conditions, that the fermion masses are given by
\beq
m_{a b} = 2 m_{3/2} \left(A_{(a d^\star b l^\star)} (A_{(i j^\star k l^\star)} f_i f_{j^\star})^{-1} f_{d^\star} f_k - \frac{f_a f_b}{f_{eff}^2} \right)
\label{fermmasssugra}
\eeq
once the goldstino direction has been rotated away.  (The goldstino direction is the zero eigenvector of this mass matrix which can clearly be seen as $f_a m_{ab} = 0$.)

Armed with Eq.(\ref{fermmasssugra}) we can apply these results to modulini from the ISS sector.  In Section \ref{ISS} we identified two quartic operators in the K\"ahler potential that may lead to modulini masses:  $\Tr[(\Phi^\dagger \Phi)^2]$ and $\Tr[(\Phi^\dagger \Phi)]^2$, and in both cases the tensor $A$ can be written in terms of the identity matrix.  First we consider just the operator $\frac{1}{\mu^2} \Tr[(\Phi^\dagger \Phi)^2]$ and diagonalise the F-terms as in Section \ref{ISS}.  In this basis one finds for the mass matrix
\beq
m_{a b, c d} = 2 m_{3/2} \left(\frac{f_a^2+f_b^2}{2 f_a f_b} \delta_{a d} \delta_{b c} - \frac{f_a f_c}{f_{eff}^2} \delta_{a b} \delta_{c d} \right)
\label{fermmass4}
\eeq
where the $f_a$ are the diagonal elements of the $h \mu_0^2$ matrix in the superpotential.  We can now split $\Phi$ into two sets of fields to study the masses.  

First, focusing on the diagonal elements, i.e. $a=b,c=d$, we find a mass matrix:
\beq
m_{a a, b b} = 2 m_{3/2} \left(\delta_{a b} - \frac{f_a f_b}{f_{eff}^2} \right)
\label{fermmass5}
\eeq
which has $N_c$ eigenvalues of $2 m_{3/2} (1,1,1,...,1,0)$.  The field with zero mass is the true goldstino field $G = f_i \Phi_{i i} / f_{eff}$ that mixes with the gravitino and is eaten leading to a gravitino of mass $m_{3/2}$.  In the presence of multiple SUSY breaking sectors this field is in general a mixture of the goldstini from all sectors and from the ISS sector we would then expect $N_c$ `goldstini' fields, $\zeta$, of mass $2 m_{3/2}$.

Now considering the off-diagonal fields, i.e. $a\neq b, c\neq d$, we find that the only non-zero terms in the mass matrix have $c=b,d=a$ and are of the form
\beq
m_{a b, b a} = 2 m_{3/2} \left(\frac{f_a^2+f_b^2}{2 f_a f_b} \right)~~.
\label{fermmass6}
\eeq
These fields in general have $m \geq 2 m_{3/2}$, with a lower limit, $m = 2 m_{3/2}$, in the case with all F-terms equal (again ignoring
warping and/or conformal sequestering).  These off-diagonal fields are the (now massive) modulini, $\chi$, which accompany the goldstini.

In summary, in the context of multiple sequestered SUSY breaking sectors, from each meta-stable ISS-type SUSY-breaking sector one
expects $N_c$ goldstini of mass $2 m_{3/2}$ and $N_c (N_c - 1)$ modulini with mass $m \geq 2 m_{3/2}$.  We emphasise that this result
is valid in the absence of extra singlet fields or other deformations of the global-SUSY-limit of the ISS sector that spoil the discrete R-symmetry outlined in Section \ref{ISSatlowE}.  Nevertheless, as explained in the Introduction, our results are expected to apply to realistic theories of SUSY-breaking in the string landscape as there is no reason to require {\it all} of the independent SUSY-breaking sectors to break their discrete R-symmetries.

\subsection{Sub-leading corrections}\label{rsymm}

There are a number of operators that might alter these results.  We first consider those we expect to arise within the ISS sector itself.  Na\"{i}vely, a cause for concern is the fact that by using the effective K\"ahler potential we are omitting higher order terms in an expansion in $f/\mu^2\lsim 1$.  We have, however, calculated the full one-loop diagram for the modulini masses, which includes these corrections to all orders and where the effects of SUGRA are included, and we find that the goldstini masses remain unaltered and the modulini masses remain bounded below by $2 m_{3/2}$, so do not change the qualitative results from the previous section.  The full results of this calculation are contained in Appendix \ref{corrections}.

Next, as discussed in \cite{ISS} and in Section \ref{ISSatlowE}, there are corrections due to the underlying microscopic theory of the form $\delta K \sim \Tr[\Phi^\dagger \cdot \Phi]^2 /| \Lambda|^2$ where $\Lambda$ is the strong coupling scale of the theory.  As highlighted in \cite{ISS} the effects from these operators are expected to be small as $|\Lambda| \gg \mu$.  Including this operator we find that the masses of the fermions are altered slightly.  In particular if we set all F-terms equal we find that for
\beq
K =  \Tr[\Phi^\dagger \cdot \Phi] - \frac{a}{|\mu|^2}  \Tr [(\Phi^\dagger \cdot \Phi)^2 ]-\frac{b}{| \Lambda|^2} \Tr [\Phi^\dagger \cdot \Phi ]^2
\eeq
and the superpotential in Eq.(\ref{lowtheoryW}) the fermion masses are
\beq
m = 2 m_{3/2} \frac{1}{1+ \frac{ b N_c  |\mu|^2}{a |\Lambda|^2}}
\eeq
As $|\Lambda| \gg |\mu|$ then unless $b\gg a$ corrections from the microscopic theory are small (though possibly phenomenologically interesting).  The sign of $b$ is unknown and so these small corrections to the goldstini and modulini masses can potentially be positive or negative.

As mentioned earlier gauge interactions also lead to an explicit breaking of the R-symmetry through the generation of the low energy superpotential \cite{ISS,Davis:1983mz,Affleck:1983mk,Intriligator:1995au,Argyres:2005cb}
\begin{equation}
W = N_c (h^{N_f} \Lambda^{-(3 N_c-2 N_f)} \det[\Pi])^{1/(N_f-N_c)} ~~.
\end{equation}
Corrections due to this term should be small, though.  First, this operator leads to a superpotential term proportional to $\Phi^{N_f/(N_f-N_c)}$; however we know that $N_f>3 (N_f-N_c)$ so the discrete R-symmetry remaining after the inclusion of this operator will forbid majorana masses for the modulini in $\Pi$ in the global SUSY limit.  On the other hand, once the theory is embedded in SUGRA, we know the R-symmetry is broken and this leads to vacuum expectation values for the scalar components of $\langle \Pi \rangle \sim m_{3/2}$ \cite{Abe:2006xp}.  As with the corrections to the K\"ahler potential we would then expect this operator to lead to masses for the modulini.  We can estimate these corrections as
\bea
\delta m & \sim & N_c h^{N_f/(N_f-N_c)} \Lambda^{-(3 N_c-2 N_f)/(N_f-N_c)} \vev{\Phi}^{N_f/(N_f-N_c)}\\
& \sim & m_{3/2} \left( \frac{m_{3/2}}{\Lambda} \right )^{(3 N_c-2 N_f)/(N_f-N_c)}
\eea
and as $\Lambda \gg \mu \gg m_{3/2}$ and also $3 N_c > 2 N_f$, these corrections should be small unless $(3 N_c-2 N_f)/(N_f-N_c) < 1$.

Other operators may arise from outside the ISS sector which lead to R-symmetry breaking and would modify these masses.  Such scenarios have been discussed in detail in \cite{Goldstini1} and thus we direct the interested reader to this work for a through discussion.  We note that if the ISS sector(s) is/are sequestered from other SUSY breaking sectors, and only couple to them via the SSM, then corrections to these masses should be at least a loop factor smaller than $m_{3/2}$.  If the ISS sector only couples to the SSM and other SUSY breaking sectors gravitationally then we expect the masses not to deviate from the calculation above.

\section{Couplings and Phenomenology}\label{pheno}

We now turn briefly to collider phenomenology.  For definiteness throughout this section, we will take as a working assumption the set-up described in detail in Refs.\cite{Goldstini1,Goldstini2}, namely, the goldstino and gravitino masses are $\geq \mathcal{O}(10 \text{GeV})$, and the SUSY-breaking scales are such that the goldstini have comparable, or greater couplings to the SSM than the gravitino.  In such a set-up all SUSY collider events will terminate in a cascade decay to the LOSP which may further decay to the goldstino or gravitino within the detector.  Such an event can lead to striking signatures at the LHC \cite{Dimopoulos:1996vz} such as monochromatic electrons or muons in the case of a selectron or smuon LOSP \cite{Nomura:2007ap}.  The lifetime of the LOSP could be determined by observing decays of stopped LOSPs within the detector \cite{Asai:2009ka,CMSLOSP} or within a proposed stopper detector \cite{Hamaguchi:2004df,Feng:2004yi,Hamaguchi:2006vu} which could be constructed after the observation of long-lived charged particles.  Further, when observing these decays it may be possible to determine the masses of the gravitino and goldstino using the methods discussed in \cite{Cheng:2007xv,Cho:2007qv,Kitano:2006gv}.  Therefore, under these assumptions, it may be possible to measure the gravitino and goldstino masses and couplings to the SSM LOSP and we will take this to be the case throughout the remainder of this work.

\subsection{Couplings in the warped/conformally-sequestered case}

Let us begin with a few brief remarks on the coupling of warped and sequestered goldstini to the Standard Model. The couplings of the goldstini to Standard Model fields in this case come from interactions of the form
\beq
\lang \supset \sum_i \frac{1}{\Lambda_i^2} \int d^4 \theta X_i^\dag X_i \Phi^\dag \Phi ~~.
\label{coupling}
\eeq
In the case of conformal sequestering (the situation is analogous for warping), large anomalous dimensions associated with the operator $X_i$ lead to a suppression of the above operator at scales $E < \Lambda_i$ of order $(E/\Lambda_i)^{2 \gamma_i}$ (if the operator $X_i^\dag X_i$ corresponds to a conserved current there is no suppression).  Assuming the exit from the conformal fixed point is controlled by SUSY breaking, this amounts to a suppression of order $(\sqrt{f_{X_i}}/\Lambda_i)^{2 \gamma_i}$ in the infrared.  Ultimately, this suppression affects both the goldstino-SSM couplings as well as the contributions of this sector to SSM soft masses, such that the infrared interactions are still of the form
\beq
\lang \supset  \sum_i \frac{1}{\Lambda_i^2} \left( \frac{f_{X_i}}{\Lambda_i^2} \right)^{\gamma_{i}} \int d^4 \theta X_i^\dag X_i \Phi^\dag \Phi = \sum_{i,a} \frac{\tilde m_i^2 V_{ia}}{f_i} \zeta_a \psi \phi^\dag ~~.
\label{coupling2}
\eeq
Here the conformal suppression is simply absorbed into the soft mass $\tilde m_i$, which therefore may be significantly smaller than naive expectations.  The principle effect of this is to further suppress the contributions of sequestered or warped sectors to both SSM soft masses and the relevant SSM-goldstino couplings. However, for {\it fixed} TeV-scale soft masses, the couplings of such goldstini to the SSM are still significantly stronger than gravitational. The couplings of, e.g., derivatively-coupled modulini in a warped/sequestered sector are suppressed relative to this by the usual factor $E / \tilde m$ at energy scale $E$, but not by additional factors from warping. 

As mentioned in the Introduction, $V_{ia}$ is the rotation matrix that diagonalises the goldstini mass matrix:
\beq
m_{ij} = 2 m_{3/2} \left(\delta_{ij} - \frac{f_i f_j}{f_{eff}^2}\right)~~,
\eeq
where the $\zeta_a$ are the $N-1$ goldstini mass eigenstates and the true goldstino that forms the longitudinal component of the
gravitino is the $N^{th}$ eigenstate with zero mass in this basis.
We see that as $\sum_i f_i V_{i,{a \neq N}}=0$ then if SUSY breaking from all sectors is communicated in an identical way, i.e.\ all $\Lambda_i$ are equal, then the goldstini couplings to the SSM would be zero, and we would only interact with the true goldstino that forms the longitudinal mode of the gravitino.   However, it is a reasonable assumption that in general not all $\Lambda_i$ are equal, and if even one of these effective mediation scales is different then couplings to all goldstini are generated.

\subsection{Distinguishing ISS SUSY-breaking}

An important question is how we could possibly distinguish if we were coupled to $N_c$ goldstini from one hidden ISS sector, multiple goldstini from many different sectors or just one goldstino with a different effective SUSY breaking scale.  We will see that making this distinction is in principle possible, however we will first consider some moral differences between these scenarios.  First of all for a hidden ISS sector one would expect a larger coupling of goldstini to sfermion-fermion pairs than gaugino-gauge boson pairs.  This would imply a hidden SUSY breaking sector that preserves an R-symmetry.  Secondly the F-terms of a hidden ISS sector should be roughly the same magnitude whereas there is no a priori reason to expect the SUSY breaking scale in multiple sequestered sectors to be similar.  Finally the SUSY breaking F-terms of a hidden ISS sector would be mediated in a similar way, and therefore couplings to goldstini arising within a single ISS sector should be of the same order of magnitude.

We illustrate the possibility of making this distinction with the example given in the Introduction of two sequestered SUSY breaking sectors which couple to the SSM differently as illustrated in Figure \ref{fig:example}. 
Considering the decay of a scalar LOSP to the goldstini via the Goldstino Portal, if we ignore details of phase-space factors, the partial width for this process is:
\bea
\Gamma_{\phi^\dagger \to \zeta \psi } & \simeq & \frac{1}{16 \pi} m_\phi \sum_{a=1}^{N-1} |C_a|^2
\eea
where $C_a$ is the dimensionless coupling of the goldstini to $\phi$ and $\psi$ given in Eq.(\ref{coupling2}).  As detailed in Appendix \ref{widthcalc}, if the sfermion masses are generated through a K\"ahler potential term of the form
\beq
K_{soft} =  \frac{\Tr [\Phi^\dagger \cdot \Phi] \phi^\dagger \phi}{x \Lambda^2}
\label{portal1}
\eeq
where $\phi$ is an MSSM field, we find that the respective decay widths to goldstini and the gravitino are:
\bea
\Gamma_{\phi^\dagger \to \zeta \psi } & \simeq & \frac{m_\phi}{16 \pi} \left( \frac{(x-1) f}{x\Lambda^2} \right)^2 \frac{ N_c y^2}{1+ N_c y^2}\\
\Gamma_{\phi^\dagger \to G \psi } & \simeq & \frac{m_\phi}{16 \pi} \left(\frac{f}{x\Lambda^2} \right)^2 \frac{(x+ N_c y^2)^2}{1+ N_c y^2}
\eea
As expected we see that in the limit $x \to 1$ the decay channel to goldstini vanishes and we recover the usual decay width to the gravitino.  More importantly however we see that $N_c$ from the ISS sector always appears in combination with $y^2$ which parameterises the overall scale of the SUSY breaking in the ISS sector.  Therefore in this scenario with Goldstino Portal couplings alone one could not distinguish the $N_c$ goldstini in an ISS sector from a SUSY breaking sector with one goldstino and a higher SUSY breaking scale.

\begin{figure}[]
\centering
\includegraphics[height=2.5in]{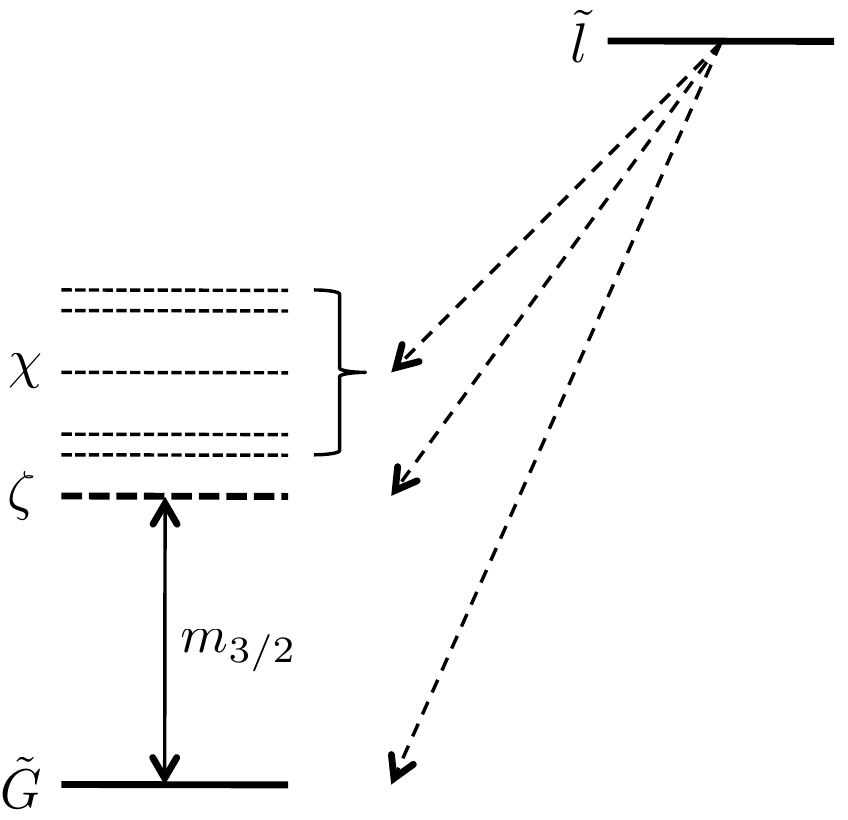}\caption{LOSP decays to the gravitino, goldstini and modulini of a hidden ISS sector.  The modulini masses are bounded below by $2 m_{3/2}$ and the observation of such a decay pattern would provide strong support for the physical realisation of the ISS mechanism of SUSY breaking.}
\label{fig:decay}
\end{figure}
However, there is in principle no reason for the couplings to be of the form in Eq.(\ref{portal1}) and the more general coupling
\beq
K_{soft} =  \frac{B_{ijkl} \Phi^\dagger_{ij} \Phi_{kl} \phi^\dagger \phi}{x \Lambda^2}
\eeq \label{portal2}
where $B$ takes some unknown values, allows not only the goldstini to couple to the SSM fields but the off-diagonal modulini fields also couple with similar strength.  This arises through the non-zero F-terms of the SUSY breaking fields.\footnote{We note that the coupling of hidden sector SUSY-preserving fields to the SSM through their interactions with the SUSY breaking fields can therefore be of the same strength as goldstino couplings to the SSM and this could have phenomenological consequences for other models of SUSY breaking.  This mechanism of SUSY-preserving fields `hitching' through the Goldstino Portal could be useful in scenarios where small renormalizable couplings to gauge-invariant combinations of SSM fields are desired.}  This coupling then allows for a `smoking gun' collider signature of a hidden ISS sector as LOSP decays will occur to the goldstini and modulini of the hidden ISS sector.  If not all F-terms are identical then the modulini masses are all greater than $2 m_{3/2}$ (in the absence of warping or conformal sequestering), but bounded below by this value and the long-lived LOSP decay spectrum would be observed to be of the form depicted in Figure \ref{fig:decay}.  This decay pattern would give strong support for the ISS mechanism of SUSY breaking if observed and the number of colours in the hidden ISS sector could, in principle, be deduced from the number of decay lines.

This signature is distinct from, say, a LOSP decaying to many gravitino-mass-scale moduli or modulini.  This is because the couplings to the ISS sector particles are not simply Planck-scale but depend on a combination of the SUSY-breaking scale in the ISS sector and the messenger scale, which needn't necessarily bear any relation to the Planck scale.  Therefore, although one would generically expect many $O(m_{3/2})$ mass particles in top down constructions such particles would not typically lead to the LOSP decay signatures as could arise from a hidden ISS sector.

\section{Conclusions}

As the LHC begins to make inroads in the exploration of the electroweak scale, it is timely to consider what experimental indications may be found regarding physics at much higher scales. Certainly the discovery of Standard Model superpartners would be a great breakthrough in itself, but there also exists the potential to learn much more about the mechanism of supersymmetry breaking and its communication to our sector. The observation of LOSP decays to a gravitino would tell us about the quantum nature of gravity \cite{Buchmuller:2004rq}, while the goldstini proposal \cite{Goldstini1} shows both that such an observation could be consistent with a standard cosmology \cite{Goldstini2} and that the observation of LOSP decays to additional goldstini would imply the existence of other sequestered SUSY breaking sectors.  As we have argued, the existence of such SUSY breaking sectors is a natural consequence of compactification on a topologically complex manifold.

Considering the consequences of multiple sequestered SUSY breaking sectors in light of experimental constraints (e.g., FCNCs) and theoretical considerations (e.g., the structure of calculable models of dynamical SUSY breaking) leads to a surprisingly rich spectrum of fields whose masses range from $0 - 2 \, m_{3/2}$ and whose interactions with Standard Model fields may be much stronger than the naive mediation scales suggest. The presence of such goldstini and modulini spanning a range of masses would significantly alter conventional supersymmetric phenomenology. Moreover, measuring their masses and couplings at the LHC would lend insight into not merely the mechanism of supersymmetry breaking, but also the existence and dynamics of additional sectors coupled to the Standard Model through the Goldstini Portal. In this fashion, various features of ultraviolet physics -- warping, conformal dynamics, metastable supersymmetry breaking -- may become evident in the infrared via the mass spectrum and interactions of goldstini and modulini.

\section*{Acknowledgements}
We gratefully thank Joe Conlon, Rhys Davies, Daniel Green and Lawrence Hall for stimulating and enjoyable discussions.  NC would like to acknowledge the Dalitz Institute for Fundamental Physics and the Department of Theoretical Physics, Oxford University for hospitality during the completion of this work, while JMR and MM would like to thank the particle theory groups at both UC Berkeley and Stanford University for hospitality during the inception of this work.  NC is supported by the NSF GRFP and the Stanford Institute for Theoretical Physics under NSF Grant 0756174. MM is supported by an STFC Postgraduate Studentship.  JMR and MM also acknowledge support by the EU Marie Curie Network ÒUniverseNetÓ (HPRN-CT-2006-035863), and JMR by a Royal Society Wolfson Merit Award. 

\appendix
\makeatletter
\def\@seccntformat#1{\csname Pref@#1\endcsname \csname the#1\endcsname\quad}
\def\Pref@section{Appendix~}
\makeatother

\section{Modulini masses from Supergravity}\label{SUGRAmass}
In order to calculate SUGRA effects on the ISS model we study a slightly more general case of the theory detailed in Section \ref{ISS}.  We start with superfields $X_i$ with superpotential
\begin{equation}
W = W_0 + f_a X_a
\end{equation}
and K\"ahler potential
\begin{equation}
K = X_a X^{\dagger}_{a^\star} + \frac{1}{|\mu|^2} A_{a b^\star c d^\star} X_a X^{\dagger}_{b^\star} X_c X^{\dagger}_{d^\star}~~,
\end{equation}
from which we may define the modified K\"ahler potential
\begin{equation}
G = \frac{K}{M_P^2} + \log{\frac{W}{M_P^3}} + \log{\frac{W^\star}{M_P^3}}
\end{equation}
and field derivatives as $G_a = \partial_a G$, $G_{a b^\star} = \partial_{a} \partial_{b^\star} G$ with $\partial_{a} = M_P \frac{\partial}{\partial X_a}$.  For a modified K\"ahler potential of this form, then, once the goldstino direction has been rotated away, the fermion mass matrix is given as \cite{WessandBagger}:
\begin{equation}
m_{a b} = m_{3/2} \langle \nabla_a G_b + \frac{1}{3} G_a G_b \rangle
\label{fermmass}
\end{equation}
where $\nabla_a G_b = \partial_a G_b - \Gamma^c_{a b} G_c$.  The Christoffel connection, $\Gamma$, is of crucial importance as it encodes the effects of $A_{a b^\star c d^\star}$.  Now considering the leading terms in the fermion mass matrix under the assumption that $\sqrt{f} \sim \mu \ll M_P$ one finds:
\begin{equation}
m_{a b} = m_{3/2} \left(-\frac{2}{3} \frac{f_a f_b M_P^2}{W_0^2} - \frac{M_P^2}{W_0 |\mu|^2} \delta^{c d^\star} A_{(a d^\star b l^\star)} f_c \langle X^\dagger_{l^\star} \rangle\right)
\label{fermmass2}
\end{equation}
where $A_{(a d^\star b l^\star)}$ has been symmetrized over pairs of holomorphic indices.\footnote{By this we mean $A_{(a d^\star b l^\star)} = A_{a d^\star b l^\star} + A_{b d^\star a l^\star} + A_{a l^\star b d^\star} + A_{b l^\star a d^\star}$.}  At this stage it is appropriate to pause and consider the validity of this result.  Throughout we have assumed that as $\sqrt{f} \sim \mu \ll M_P$ then $\langle X \rangle \ll M_P$ and $W_0 \ll M_P^3$.  It may seem that if one takes the limit $A\rightarrow0$ then $m_{a b} \propto f_a f_b$ which is a rank one matrix with only one non-zero eigenvalue.  However taking this limit means that the scalar fields are no longer stabilised near the origin and the derivation of this result is no longer valid.  Also it would appear from this result that the fermion mass matrix depends on the parameter $\mu$; however we will see that $\langle X \rangle \sim |\mu|^2 / M_P$ and this dependence drops out.  Again, this independence only necessarily holds in the limit $\mu \ll M_P$.

Now considering the scalar potential $V = M_P^4 e^G (G_a G^a - 3)$ one finds that for vanishing cosmological constant
\begin{equation}
W_0 = M_P \sqrt{\frac{f_a f_a}{3}} = \frac{1}{\sqrt{3}} f_{eff} M_P
\label{w0}
\end{equation}
and at the minimum of the scalar potential
\begin{equation}
\langle X^\dagger_{l^\star} \rangle = - \frac{2 |\mu|^2 W_0}{M_P^2} f_k (A_{(a b^\star k l^\star)} f_a f_{b^\star})^{-1}
\label{vev}
\end{equation}
where $(M_{k l^\star})^{-1}$ is understood as the standard matrix inverse.  With these results in hand we can now write a general formula for the modulini mass matrix
\begin{equation}
m_{a b} = 2 m_{3/2} \left(A_{(a d^\star b l^\star)} (A_{(i j^\star k l^\star)} f_i f_{j^\star})^{-1} f_{d^\star} f_k - \frac{f_a f_b}{f_{eff}^2} \right) ~~.
\label{fermmass3}
\end{equation}
This equation is valid up to corrections of the order $\delta m \sim m_{3/2} |\mu|^2/M_P^2$.  The extension of this formula to one for matrix-valued fields can be found by replacing individual indices with pairs, i.e. $\{a\} \rightarrow \{a b\}$.  At first Eq.(\ref{fermmass3}) may appear rather opaque, however one important property can be observed by inspection.   As described in \cite{WessandBagger} once the goldstino direction has been rotated away one expects that $G_a m_{a b} = 0$.  This is clear from Eq.(\ref{fermmass}) when one enforces the condition of vanishing cosmological constant and that the fields are at the minimum of the potential.  At the level of Eq.(\ref{fermmass3}) one can see that this result also holds for any form of $A_{a b^\star c d^\star}$ as $f_a m_{a b} = f_b - f_b = 0$ by inspection.

\section{Masses to all orders in $f$}\label{corrections}
As described in Section \ref{ISSatlowE} the effective K\"ahler potential only includes corrections to second order in the SUSY breaking F-terms.  To include higher order corrections at the level of the K\"ahler potential would require including higher order supercovariant derivatives.  Therefore it is more straightforward to calculate the modulini masses to all orders in the F-terms by explicitly evaluating the loop diagram involving the exchange of scalar and fermionic partners of the heavy fields.  In this manner the effects of SUGRA, and consequent R-symmetry breaking, are included by allowing for a non-zero vacuum expectation value for the fields which break SUSY.  This vev can be calculated to all orders in the F-terms by including the tadpole term induced by SUGRA and calculating the scalar masses with the Coleman-Weinberg potential.  Evaluating the one-loop contribution to the modulini masses using the masses and couplings in Eq.(\ref{SuPlow}), and setting $h=1$ for convenience, one finds
\begin{equation}
m_{a b, c d} = 2 m_{3/2} \left(\frac{1}{2} \left(\frac{H(f_a)}{H(f_b)} +\frac{H(f_b)}{H(f_a)}\right) \delta_{a d} \delta_{b c} - \frac{f_a f_c}{f_{eff}^2} \delta_{a b} \delta_{c d} \right)
\end{equation}
where
\begin{equation}
H(f) = \sum_{i=1}^{N_f-N_c} h(f,\tilde{\mu}_{0_i}^2)
\end{equation}
and
\begin{equation}
h(f,\mu^2) = \frac{1}{f^2} \left(2 f \mu^2+2 f \mu^2 \log\left(\frac{\mu^4}{\mu^4-f^2} \right) + (\mu^4+f^2) \log\left(\frac{\mu^2-f}{\mu^2+f} \right) \right)
\end{equation}
Here $\tilde{\mu}_{0_i} \geq \sqrt{f}$ is the SUSY mass of the fields which have been integrated out.  One can see that all dependence on the UV cut-off has cancelled and the masses are finite.  The goldstini from the diagonal components of $\Phi$ still have mass $2 m_{3/2}$ and the modulini from the off-diagonal components have mass $\geq 2 m_{3/2}$, limiting to $2 m_{3/2}$ when the F-terms are equal, as before.  Therefore the results derived using the effective K\"ahler potential in Section \ref{ISSatlowE} are qualitatively the same as those one finds when including the F-terms to all orders.

\section{Decay widths}\label{widthcalc}
Starting with Eq.(\ref{coupling}) we derive the decay width to multiple goldstini under the assumption that all but one messenger scales are the same.  We take the first $N-1$ messenger scales equal to $\sqrt{x} \Lambda$ and the $N^{th}$ messenger scale as $\Lambda$.  Using this, the orthogonality of $V_{ia}$, the fact that $V_{N i} = f_i/f_{eff}$ and that $\sum_i f_i V_{i,{a \neq N}}=0$ we can simplify the sum over squares of the couplings:
\bea
\sum_{a=1}^{N-1} |C_a|^2 & = & \sum_{a=1}^{N-1} \left |\sum_{i=1}^{N} \frac{f_i V_{ia}}{\Lambda_i^2} \right |^2 \nonumber \\
& = & \sum_{a=1}^{N-1} \frac{(x-1)^2 f_N^2}{x^2 \Lambda^4}V_{N a} V_{a N}^{T} \nonumber \\
& = & \frac{(x-1)^2 f_N^2}{x^2 \Lambda^4} \frac{f_{eff}^2 - f_N^2}{f_{eff}^2}
\eea
Thus we have
\bea
\Gamma_{\phi^\dagger \to \zeta \psi } & \simeq & \frac{m_\phi}{16 \pi} \left( \frac{(x-1) f_N}{x\Lambda^2} \right)^2 \frac{f_{eff}^2 - f_N^2}{f_{eff}^2}
\eea
For decays to the gravitino similar steps lead to
\bea
|C_N|^2 & = & \left( \frac{f_{eff}^2 - (1-x) f_N^2}{x \Lambda^2 f_{eff}} \right)^2
\eea
and
\bea
\Gamma_{\phi^\dagger \to G \psi } & \simeq & \frac{m_\phi}{16 \pi} \left( \frac{f_{eff}^2 - (1-x) f_N^2}{x \Lambda^2 f_{eff}} \right)^2
\eea
These results make no assumptions about the relative magnitudes of the various F-terms and therefore hold if there are multiple SUSY breaking sectors and all but one mediate SUSY breaking to the SSM in the same way.  If all mediation sectors are the same this corresponds to the limit $x\to1$.

\bibliographystyle{JHEP}
\bibliography{myrefs}

\end{document}